\documentclass[aps,pra,twocolumn,showpacs,amsmath,floatfix]{revtex4-1}

\usepackage{graphicx}
\usepackage{dcolumn}
\usepackage{multirow}

\newcommand{\figurewidth}{\columnwidth}

\begin{document}

\title{Storage-ring measurement of the hyperfine-induced  $\mathbf{2s\,2p\;^3P_0\to2s^2\;^1S_0}$ transition rate in beryllium-like sulfur}

\author{S.~Schippers}
\email[]{Stefan.Schippers@physik.uni-giessen.de}
\homepage[]{\linebreak http://www.strz.uni-giessen.de/cms/iamp}
\affiliation{Institut f\"{u}r Atom- und Molek\"{u}lphysik, Justus-Liebig-Universit\"{a}t Giessen, Leihgesterner Weg 217, 35392 Giessen, Germany}

\author{D.~Bernhardt}
\affiliation{Institut f\"{u}r Atom- und Molek\"{u}lphysik, Justus-Liebig-Universit\"{a}t Giessen, Leihgesterner Weg 217, 35392 Giessen, Germany}

\author{A.~M\"{u}ller}
\affiliation{Institut f\"{u}r Atom- und Molek\"{u}lphysik, Justus-Liebig-Universit\"{a}t Giessen, Leihgesterner Weg 217, 35392 Giessen, Germany}

\author{M.~Lestinsky}
\affiliation{GSI Helmholtzzentrum f\"{u}r Schwerionenforschung, Planckstrasse 1, 64291 Darmstadt, Germany}

\author{M.~Hahn}
\affiliation{Columbia Astrophysics Laboratory, Columbia University, 550 West 120th Street, New York, NY 10027, USA}

\author{O.~Novotn\'y}
\affiliation{Columbia Astrophysics Laboratory, Columbia University, 550 West 120th Street, New York, NY 10027, USA}

\author{D.~W.~Savin}
\affiliation{Columbia Astrophysics Laboratory, Columbia University, 550 West 120th Street, New York, NY 10027, USA}

\author{M.~Grieser}
\affiliation{Max-Planck-Institut f\"{u}r Kernphysik, Saupfercheckweg 1, 69117 Heidelberg, Germany}

\author{C.~Krantz}
\affiliation{Max-Planck-Institut f\"{u}r Kernphysik, Saupfercheckweg 1, 69117 Heidelberg, Germany}

\author{R.~Repnow}
\affiliation{Max-Planck-Institut f\"{u}r Kernphysik, Saupfercheckweg 1, 69117 Heidelberg, Germany}

\author{A.~Wolf}
\affiliation{Max-Planck-Institut f\"{u}r Kernphysik, Saupfercheckweg 1, 69117 Heidelberg, Germany}

\date{\today}

\begin{abstract}
The hyperfine induced $2s\,2p\;^3P_0\to2s^2\;^1S_0$ transition rate in Be-like sulfur was measured by monitoring the decay of isotopically pure beams of $^{32}$S$^{12+}$ and $^{33}$S$^{12+}$ ions in a heavy-ion storage ring. Within the 4\% experimental uncertainty the experimental value of $0.096(4)$~s$^{-1}$ agrees  with the most recent theoretical results of Cheng et al.\ [Phys.\ Rev. A \textbf{77}, 052504 (2008)] and Andersson et al.\ [Phys.\ Rev.\ A \textbf{79}, 032501 (2009)]. Repeated experiments with different magnetic fields in the storage-ring bending magnets demonstrate that artificial quenching of the $2s\,2p\;^3P_0$ state by these magnetic fields is negligible.
\end{abstract}



\maketitle

\section{Introduction\label{sec:intro}}

\begin{figure}[ttt]
\centering{\includegraphics[width=0.5\figurewidth]{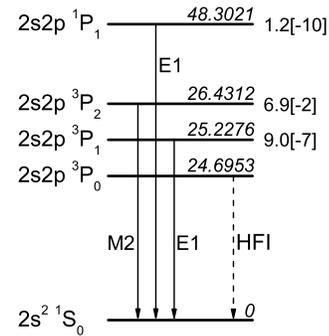}}
\caption{\label{fig:S12levels} Simplified level diagram for
beryllium-like S$^{12+}$. The level energies \cite{Ralchenko2011} are given in eV on top of the horizontal lines which are not drawn to scale.
The level lifetimes (in s) labeling the excited levels on the right side of the horizontal lines were calculated from theoretical one-photon
transition rates \cite{Landi2008} that do not account for hyperfine
effects. The one-photon transitions are labeled E1 (electric dipole), M2 (magnetic quadrupole), and HFI (hyperfine induced). Numbers in square brackets denote powers of 10. In case of
nonzero nuclear spin the hyperfine induced $^3P_0 \to {^1S_0}$ transition
rate $A_\mathrm{HFI}$ acquires a finite value.}
\end{figure}

Hyperfine quenching  in atoms and ions \cite{Johnson2011} is the shortening of excited-state lifetimes by the interaction of the electron shell with the magnetic moment of the atomic nucleus. A particularly drastic hyperfine quenching effect is observed  in alkaline-earth-like and, in general, divalent atoms and ions (with valence shell $n$) where the first excited level above the ground state is the $ns\,np\;^3P_0$ state. (\nocite{Ralchenko2011,Landi2008}Fig.\ \ref{fig:S12levels}). A total electronic angular momentum of $J=0$ for this level makes a single-photon decay to the $(ns)^2\;^1S_0$ ground state impossible. The $ns\,np\;^3P_0$ states are thus extremely long-lived considering the fact that the lifetimes associated with the two-photon E1M1 transition can be up to several months \cite{Schmieder1973a,Laughlin1980a}.

However, a nucleus with nonzero spin induces a mixing of the $ns\,np\;^3P_0$ state with its neighboring $ns\,np\;^3P_1$ state via the hyperfine interaction. This drastically increases the $ns\,np\;^3P_0\to (ns)^2\;^1S_0$ transition rate.  Such hyperfine-induced (HFI) decay rates have been calculated theoretically for beryllium-like ($n$=2) \cite{Marques1993,Brage1998a,Cheng2008a,Andersson2009,Li2010}, magnesium-like ($n$=3) \cite{Marques1993a,Brage1998a,Kang2009,Andersson2010}, and zinc-like ($n$=4) ions \cite{Liu2006a,Marques2007a,Chen2011}. Calculations have been carried out also for divalent heavier atoms \cite{Porsev2004a,Santra2004a} and singly charged ions \cite{Joensson2007}, where the long isotope-dependent lifetimes are attractive systems for obtaining ultraprecise optical frequency standards \cite{Ludlow2008,Rosenband2008}. A comprehensive review of the field, which also covers other classes of atomic systems such as He-like ions, has been published recently \cite{Johnson2011}.

So far, very few experimental measurements of hyperfine-induced (HFI) decay rates for $ns\,np\;^3P_0$ states in divalent ions have been performed.
Experiments with singly charged In$^+$ and Al$^+$ ions \cite{Becker2001a,Rosenband2007a} have used radio-frequency traps. The rates for the
atomic-clock transitions $5s5p\;^3P_0 \to 5s^2\;^1S_0$ in $^{115}$In$^+$ and $3s3p\;^3P_0 \to 3s^2\;^1S_0$ in $^{27}$Al$^+$ have been measured using optical methods with uncertainties \footnote{Throughout the paper uncertainties are quoted at a one sigma level.} of  $\sim4\%$ and $\sim7\%$, respectively. For Be-like N$^{5+}$ the HFI rate has been extracted from
observations of a planetary nebula \cite{Brage2002a} with an uncertainty of 33\%. In astrophysics HFI transitions can be used to infer isotopic abundance ratios which, in turn, provide insight into stellar nucleosynthesis \cite{Rubin2004a}.

\begin{table}[t]
\caption{\label{tab:ti18comp}Comparison of theoretical values for the hyperfine induced $2s\,2p\; ^3P_0 \to 2s^2\; ^1S_0$ transition rate in Be-like Ti$^{18+}$ with the only published experimental result.}
\begin{ruledtabular}
\begin{tabular*}{\columnwidth}{@{\extracolsep{\fill}}ccccc}
\multirow{2}{*}{Year} & \multicolumn{2}{c}{$A_\mathrm{HFI}$ (s$^{-1}$)} & Deviation from & \multirow{2}{*}{Ref.} \\
     & Experiment & Theory & experiment & \\
     \hline\rule[0mm]{0mm}{4mm}%
 1993 &           & 0.3556           & -37\%           & \cite{Marques1993} \\
 2007 & 0.56(3)   &                  &                 &  \cite{Schippers2007a}\\
 2008 &           & 0.6727           & \phantom{-}20\% & \cite{Cheng2008a}\\
 2009 &           & 0.6774           & \phantom{-}21\% &  \cite{Andersson2009}\\
 2010 &           & 0.661\phantom{4} & \phantom{-}18\% & \cite{Li2010}
\end{tabular*}
\end{ruledtabular}
\end{table}

A laboratory measurement \cite{Schippers2007a} of a HFI transition rate in a multiply charged beryllium-like ion was carried out with Ti$^{18+}$ employing the same storage-ring technique as in the present work. The experimental value for the HFI $2s\,2p\;^3P_0\to2s^2\;^1S_0$ transition rate is $0.56(3)$~s$^{-1}$. The only theoretical value \cite{Marques1993}, that was available at the time of the experiment, was significantly smaller (Tab.\ \ref{tab:ti18comp}). This first laboratory measurement of a HFI induced transition rate in a Be-like ion has triggered new theoretical calculations not only for Be-like ions \cite{Cheng2008a,Andersson2009,Li2010} but also for Mg-like ions \cite{Kang2009,Andersson2010} and Zn-like ions \cite{Marques2007a}. Table \ref{tab:ti18comp} compares the experimental HFI $2s\,2p\;^3P_0\to 2s^2\;^1S_0$ transition rate for Be-like $^{47}$Ti$^{18+}$ with all presently available theoretical values. The new theoretical results agree slightly better with the experimental finding than does the older theory. In contrast to the early value, which is smaller than the experimental rate by $37\%$,  the new ones are larger by up to $21\%$.

The main technical difference between the old and the new calculations is the treatment of the atomic structure. Be-like ions have a rather compact electron shell where correlation effects are particularly strong. It turns out that the calculated values for the $2s\,2p\to 2s^2$ transition rates depend sensitively on the theoretical description of the atomic states, i.e., on the size of the configuration space considered and on the inclusion of relativistic and QED effects \cite{Andersson2009}. HFI transition rates are thus excellent probes of atomic correlations.

At present, the origin for the remaining significant deviation by 18\% from the experimental result is unclear. It has been suspected that the magnetic fields in the storage-ring bending-magnets lead to an additional quenching of the $2s\;2p\;^3P_0$ state. However, a recent theoretical investigation \cite{Li2011} concluded that this effect is negligible under the conditions of the Ti$^{18+}$ experiment.

Here we present storage ring measurements of the HFI $2s\,2p\; ^3P_0 \to 2s^2\; ^1S_0$ transition rate in Be-like $^{33}$S$^{12+}$ using the same experimental technique as before \cite{Schippers2007a}. The aim of the present investigation is to enlarge the experimental data base and to possibly shed some light onto the origin of the discrepancies discussed above. In particular, measurements have been carried out at different magnetic field strengths in order to experimentally investigate the magnetic quenching effect.

\section{Experiment}\label{sec:exp}

\begin{table}[b]
\caption{\label{tab:prop}Some relevant properties of $^{32}$S$^{12+}$ and $^{33}$S$^{12+}$ ions. Listed are the mass number $A$, mass $m$ \cite[][in atomic mass units]{Audi2003}, nuclear spin $I$, nuclear magnetic moment $\mu$ \cite [][in units of the nuclear magneton $\mu_N$]{Stone2005a}, natural abundance $f$ \cite{Boehlke2005a}, and the lifetime $\tau_{E1M1}$ \protect\citep[][HFI quenching not considered]{Schmieder1973a}, due to the $2s\,2p\;^3P_0\to2s^2\;^1S_0$ E1M1 two-photon decay. The uncertainties of $m$, $\mu$, and $f$ are smaller than the respective numerical precision of the tabulated values. }
\begin{ruledtabular}
\begin{tabular}{c@{\extracolsep{\fill}}c@{\extracolsep{\fill}}c@{\extracolsep{\fill}}c@{\extracolsep{\fill}}r@{\extracolsep{\fill}}rcrc}
  $A$ &$m$ (u) & $I$ & $\mu$ ($\mu_N$) & $f$ (\%) & $\tau_{E1M1}$ (s)\\
\hline\rule[0mm]{0mm}{4mm}%
 32 & 31.9655 &  0    &     0     & 95.041    & 6.3$\times$10$^6$ \\
 33 & 32.9649 & 3/2   & 0.64382   &  0.749    & 6.3$\times$10$^6$ \\
\end{tabular}
\end{ruledtabular}
\end{table}

The experiment was performed using the accelerator and storage-ring facilities of the  Max-Planck-Institut f\"{u}r Kernphysik (MPIK) in Heidelberg, Germany.
Negatively charged ions from a sputter ionization source were injected into the MPIK tandem accelerator.  In order to produce $^{33}$S beams, isotopically enriched FeS samples with a relative $^{33}$S  abundance of 99.8\% were used in the ion source. The FeS samples for the production of $^{32}$S beams contained sulfur with a natural isotope distribution (Tab.~\ref{tab:prop}).

Multiply charged sulfur ions were produced by electron stripping in a gas stripper located on the high-voltage terminal of the MPIK tandem accelerator and, subsequently at a higher ion energy, in a 5~$\mu$g~cm$^{-2}$ carbon foil located behind the exit of the tandem accelerator. Because of the statistical nature of the electron stripping process, ions were produced in a range of charge states. The stripping also leads to internal excitation of the resulting ions. Most of the excited states decay rapidly to the ground state and to long-lived excited states such as the S$^{12+}$($2s\,2p\;^3P$) states (Fig.~\ref{fig:S12levels}) which are of particular interest in the present work. The centroid of the charge state distribution depends on the ion energy which was chosen to be about 56~MeV at the carbon foil. At this energy, charge state 12+ was produced with nearly maximum efficiency. This charge state was selected for further beam transport by passing the beam through a dipole magnet which dispersed the various ion-beam components according to their mass-to-charge ratios.

\begin{figure}[t]
\includegraphics[width=\figurewidth]{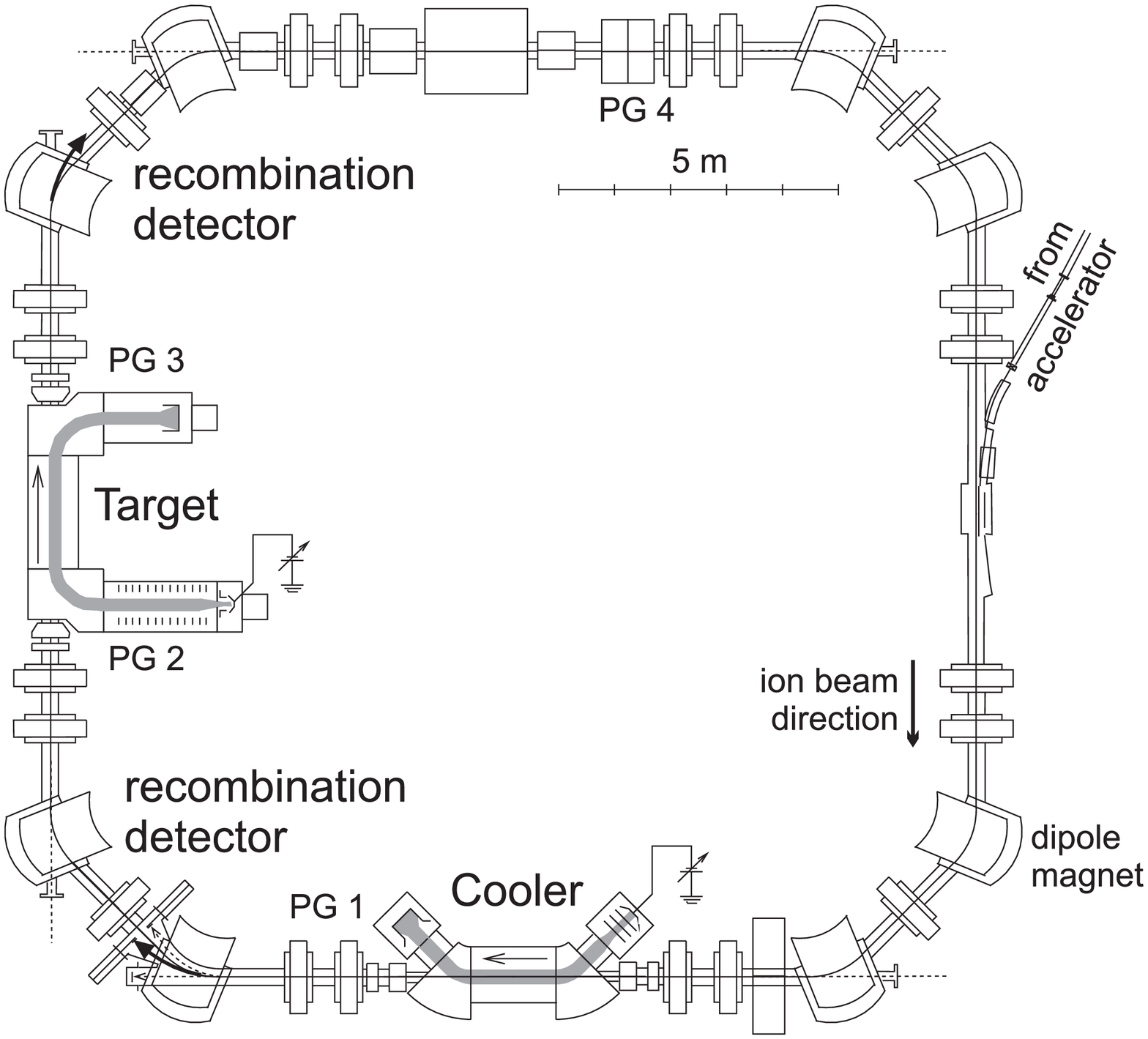}%
\caption{\label{fig:TSR}Sketch of the heavy-ion storage-ring TSR. The 8 bending dipole magnets with 1.15~m bending radius together cover 13\% of the closed orbit of the ions which has a circumference of 55.4~m. Both the electron target (Target) and the electron cooler (Cooler) were used for the present electron-ion recombination measurement. The locations of the recombination detectors are indicated by the thick curved arrows in the top left and bottom left corners of the ring which represent the paths of recombined ions. The labels PG1 -- PG4 denote the approximate locations of the pressure gauges that
were monitored by the data acquisition system.}
\end{figure}

Isotopically pure pulses of S$^{12+}$ ions were then injected into the heavy-ion storage-ring TSR. The dipole bending magnets of the storage ring were adjusted such that the S$^{12+}$ ions were circulating on a closed orbit. Stored S$^{12+}$ ion currents were up to 35~$\mu$A. The TSR storage ring \cite{Wolf2006c} is equipped with two electron-beam arrangements, which are referred to as the \lq\lq Cooler\rq\rq and the \lq\lq Target\rq\rq (Fig.~\ref{fig:TSR}) and which can be used for electron cooling and for electron-ion collision studies. For the present measurements various modes of operation were employed where either the Target or the Cooler was used for electron cooling \cite{Poth1990} and also as an electron target for electron-ion collision experiments. Details will be given below.

The electron-ion collisions that occur in the merged-beams regions can lead to formation of S$^{11+}$ ions. These recombination products are less strongly deflected in the TSR dipole magnets than the more highly charged primary ions. This enables the recombined S$^{11+}$ ions to be easily counted with single-particle detectors which are positioned such that they intercept the path of the S$^{11+}$ ions (Fig.~\ref{fig:TSR}). A channeltron-based detector \cite{Rinn1982} was used behind the Cooler and a scintillation detector \cite{Lestinsky2009} was employed behind the Target.
Because the ions moved with high velocities of 5--10\% of the speed of light the detection efficiency was practically 100\% for both detectors.

\subsection{Dielectronic recombination spectra}

\begin{figure}[t]
\centering{\includegraphics[width=0.8\figurewidth]{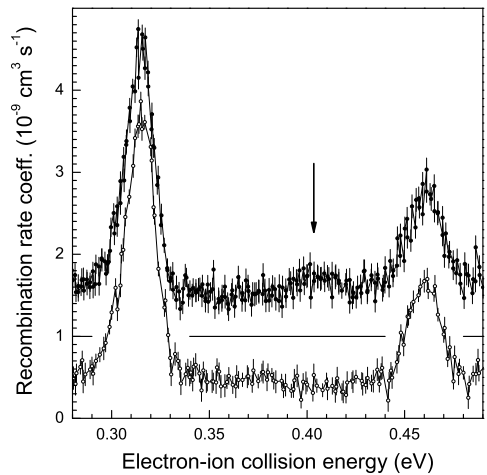}}%
\caption{\label{fig:S12DR} Measured DR spectra of $^{32}$S$^{12+}$ (full symbols, top curve, offset by $+10^{-9}$~cm$^{3}$~s$^{-1}$) and $^{33}$S$^{12+}$ (open symbols, bottom curve).
For these measurements the Target was used as electron target and the Cooler was used to continuously cool the ion beam during recombination data taking. The vertical arrow marks the position of a resonance which is associated with primary ions in the $2s\,2p\;^3P_0$ state (see discussion in the text).}
\end{figure}

After injection of ion pulses into the TSR and subsequent electron-cooling of the stored ion beam, the recombined S$^{11+}$ ion signal was recorded as a function of electron-ion collision energy for both $^{32}$S$^{12+}$ and $^{33}$S$^{12+}$ ions using well established procedures (e.g., \cite{Lestinsky2009,Schmidt2007b,Schippers2001c}, and references therein). Further details and comprehensive results of the S$^{12+}$ recombination measurements will be published elsewhere, since only a very narrow range of electron-ion collision energies is of interest for the present study.

Figure \ref{fig:S12DR} shows recombination spectra of $^{32}$S$^{12+}$ and $^{33}$S$^{12+}$ ions in the 0.28--0.49~eV energy range. In both spectra two strong DR resonances appear at 0.315 and 0.460~eV. These belong to the $(2s\,2p\;^3P)\,9l$ manifold of resonances associated with $2s\to 2p$ excitation of S$^{12+}$($2s^2$) ground-state ions. The much weaker resonance at 0.4~eV is only visible in the $^{32}$S$^{12+}$ spectrum. It is associated with the excitation of ions which are initially in the practically infinitely long-lived $2s\,2p\;^3P_0$ metastable excited state.  The resonance appears rather weak since only a relatively small fraction of Be-like ions were in the $2s\,2p\;^3P_0$ state. For isoelectronic Ti$^{18+}$ \cite{Schippers2007b} and Fe$^{22+}$ \cite{Savin2006a} ions stored in TSR, metastable fractions of 5\% and 7\%, respectively, were inferred. A similar value can be assumed here. The lifetimes of the $2s\,2p\;^3P_1$ and $2s\,2p\;^3P_2$ states (Fig.~\ref{fig:S12levels}) are orders of magnitude shorter than the 1.8~s long time interval that was used for electron cooling after each injection of new ions into the storage ring and before data taking was started. Therefore, it can be safely assumed that the fractions of primary ions in the $2s\,2p\;^3P_1$ and $2s\,2p\;^3P_2$ states were negligibly small  during all measurements.

For the $^{33}$S isotope, the $^3P_0$ resonance at 0.4~eV does not contribute to the DR spectrum because the $2s\,2p\;^3P_0$ state is quenched by hyperfine interaction. There is sufficient time for its decay since the ions are stored for much longer than the longest predicted S$^{12+}$($2s\,2p\;^3P_0$) HFI lifetime of about 28~s \cite{Marques1993,Cheng2008a,Andersson2009,Li2010}.

\subsection{Resonance-decay curves}

\begin{table}[b]
\caption{\label{tab:exppar}Some of the experimental parameters that were used during the decay-curve measurements: Atomic mass number ($A$), magnetic field in the TSR-dipoles ($B_\mathrm{dip}$), ion velocity ($v_\mathrm{ion}$),
relativistic $\gamma$-factor of the ions, and electron density ($n_e$) in the Cooler at 0.4~eV electron-ion collision energy.}
\begin{ruledtabular}
\begin{tabular}{c  c c c c c}
   data &  \multirow{2}{*}{$A$} &  $B_\mathrm{dip}$  &   $v_\mathrm{ion}$   & \multirow{2}{*}{$\gamma$}  &       $n_e$         \\
   set  &                       &   (T)              & ($10^9$~cm~s$^{-1}$) &                            &  ($10^7$~cm$^{-3}$) \\
 \hline\rule[0mm]{0mm}{4mm}
\multirow{2}{*}{{\normalsize A}} & 32 & 0.445 & 1.848 & 1.0019 &  1.93 \\
                                 & 33 & 0.442 & 1.783 & 1.0018 &  1.81 \\
 \hline\rule[0mm]{0mm}{4mm}
\multirow{2}{*}{{\normalsize B}} & 32 & 0.443 & 1.842 & 1.0019 &  1.91 \\
                                 & 33 & 0.445 & 1.792 & 1.0018 &  1.81 \\
 \hline\rule[0mm]{0mm}{4mm}
\multirow{2}{*}{{\normalsize C}} & 32 & 0.870 & 3.598 & 1.0073 &  2.64 \\
                                 & 33 & 0.883 & 3.543 & 1.0071 &  2.46 \\
\end{tabular}
\end{ruledtabular}
\end{table}

In order to obtain an experimental value for the HFI lifetime of the $2s\,2p\;^3P_0$ state, the decay of the 0.4~eV resonance was monitored as a function of storage time. To this end only one ion pulse was injected into the TSR storage ring. After a short initial cooling period the electron-ion collision energy in the Cooler was set to 0.4~eV and the recombination signal from the $^3P_0$ DR resonance was recorded for up to $\sim$220~s. Then, the remaining ions were kicked out of the TSR and a new ion pulse was injected. This sequence was repeated for several hours in order to reduce statistical uncertainties to a level as low as achievable within the limited amount of available beam time.

Because of a small vacuum leak in the recombination detector at the Target, this detector could not be used without compromising the residual gas pressure in the TSR and, consequently, considerably shortening the beam lifetime. Therefore, for all resonance-decay curve measurements the Cooler was used as an electron target.

As mentioned above, a change of the isotope required the change of the sample in the ion source. Moreover, the accelerator settings had to be slightly retuned. The entire procedure for changing the isotope took 2-4 hours. Therefore, isotope changes were performed only every other day.

\begin{figure}[t]
\includegraphics[width=\figurewidth]{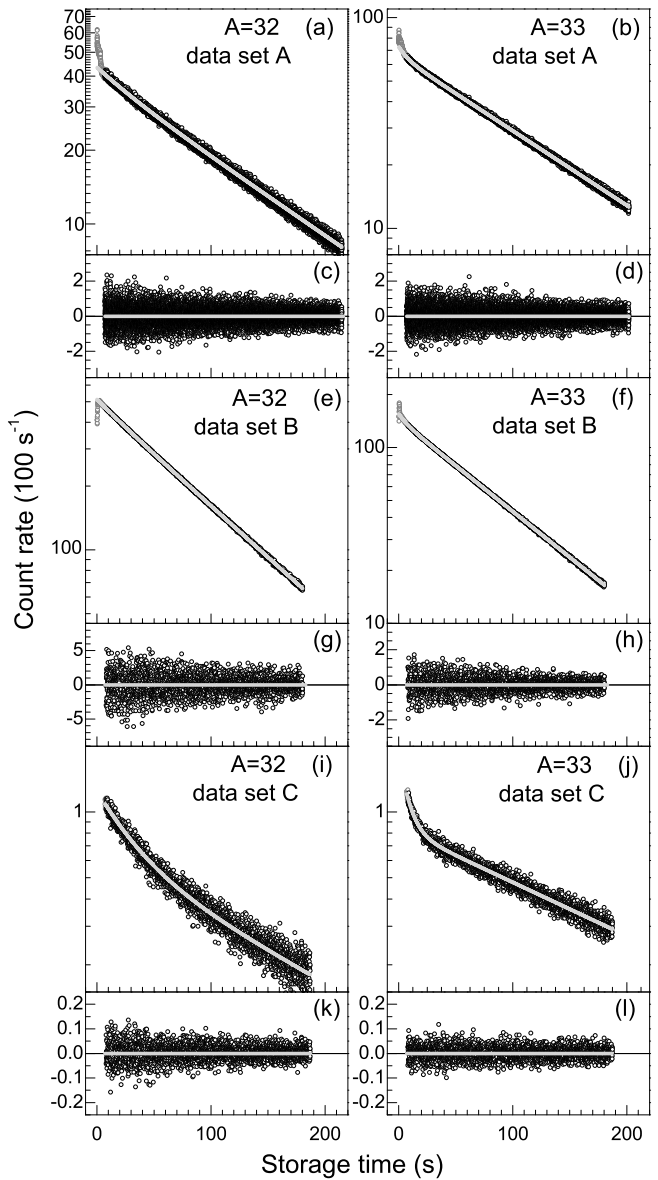}
\caption{\label{fig:allfits} Large panels: Measured resonance-decay curves (symbols) and fits of Eq.~\ref{eq:Rate2} to the data (full curves). Small panels: Residuals from the fits.}
\end{figure}

Different data sets (labeled A, B, and C) were obtained under quite different experimental conditions. Data set A was measured in October 2010. Data sets B and C were collected in April 2011. Some experimental parameters are given in Tab.~\ref{tab:exppar} and further details are discussed below. All measured beam-decay curves are displayed in Fig.~\ref{fig:allfits}.

\subsubsection{General properties}

In the decay curves in Fig.~\ref{fig:allfits}, a few general properties can be observed:

(a) Substantial differences are seen between the three pairs of curves for the data sets A--C, respectively. This arises because the electron-induced count rates are superimposed on backgrounds of varying sizes, which originate from from charge changing collisions of the ions with the residual gas in the TSR. In sets A and B, the relative contribution of the background to the measured signal is much larger than in C. Therefore, the A and B curves are dominated by the collision-induced ion beam loss. The electron-induced signal here represents a small contribution only, though it still can be analyzed as discussed below. In contrast, the electron-induced signal dominates for case C.

(b) In Fig.~\ref{fig:allfits}i of data set C, the electron-induced destruction of primarily the metastable S$^{12+}(2s\,2p\;^3P_0)$ ions leads to a rather fast beam loss during the first $\sim$30~s. Later, only the collision-induced signal from the ground-state beam, with longer lifetime, is left.

(c) In Fig.~\ref{fig:allfits}j of data set C, an even faster initial decay is clearly seen. This is caused by the additional, HFI radiative decay of the metastable state, leading to a corresponding sharp decrease of the electron-induced signal, before the collision-induced background takes over.   Hence, in the comparison of Figs.~\ref{fig:allfits}i and \ref{fig:allfits}j the HFI quenching is very clearly visible.

(d) In the curves for data sets A and B, this behavior occurs much less prominently because of the large background. However, small fast decay components from the HFI decay can be seen also in Figs.~\ref{fig:allfits}b and \ref{fig:allfits}d, comparing to the corresponding set for the even-$A$ isotope.

Using the model given in Sec.~\ref{sec:time}, time constants for the various decay modes can be extracted from all three data sets. First, though, we separately specify the measuring conditions of the sets.

\subsubsection{Data set A}

Data set A (Figs.~\ref{fig:allfits}a and b) was taken with the ions stored at their injection energy of about 56~MeV. More precise values for the ion velocity are given in Tab.~\ref{tab:exppar}. The Cooler was used as a recombination target with the
electron-ion collision energy set to 0.4~eV. The Target was used for continuous electron cooling of the ion beam.

\begin{figure}
\includegraphics[width=\figurewidth]{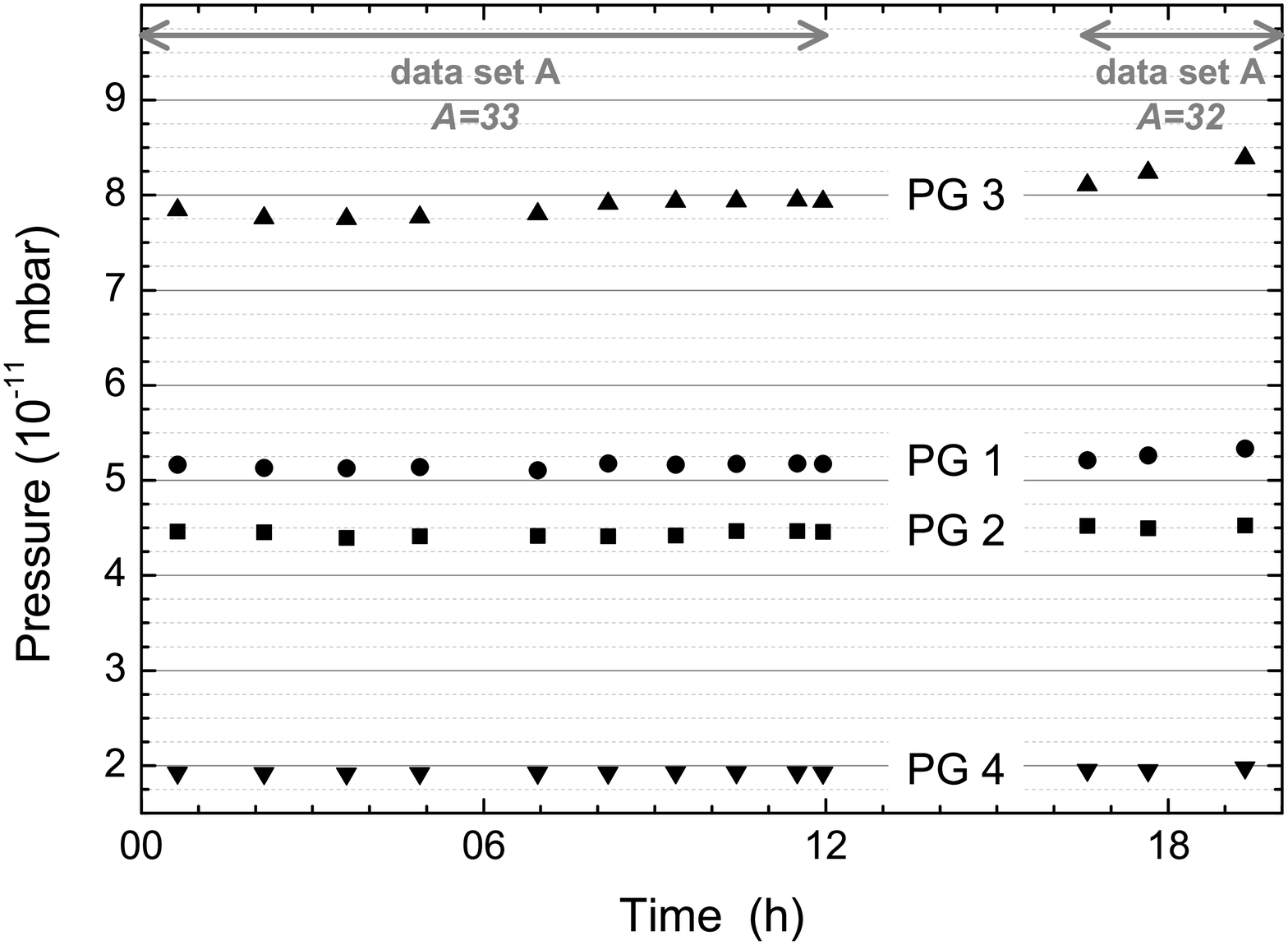}%
\caption{\label{fig:pressures} Vacuum pressures at different locations (labels PG1--4 in Fig.~\ref{fig:TSR}) in the TSR storage ring  recorded during the lifetime measurements which contributed to data set A with a single photocathode. Time zero is October 1st, 2010 at 16:14. During the gap between 12 and 16 h the isotope was changed from $^{33}$S to $^{32}$S. The pressures remained nearly constant over the entire measurement time of data set A.  It should be noted that the absolute pressure values bear rather large (unknown) uncertainties.}
\end{figure}

This mode of operation provided stable reproducible conditions only for up to 24 hours. After that the photocathode \cite{Orlov2005a} of the Target routinely had to be replaced by a freshly prepared one. It turned out that the emission characteristics of the various photocathodes used were significantly different, causing different Target electron beam profiles, and the residual gas pressures in the Target region of the TSR changed after a new cathode was installed at the Target. Since the altered conditions influenced the ion-beam lifetime considerably, measurements before and after a cathode change could not be combined into one data set. Nevertheless, we were able to measure one set of decay curves  under stable conditions (Fig.~\ref{fig:pressures}) with one photocathode for both isotopes.

\subsubsection{Data set B}

In order to avoid complications due to Target cathode changes, the Target was switched off during the measurements of data sets B and C. The Cooler was used for both electron cooling and as recombination target. The cooler is equipped with a thermionic cathode which can easily be operated with stable performance for several days. The electron energy was switched alternatingly between cooling energy, i.e., zero electron-ion collision energy, and an electron-ion collision energy of 0.4 eV. Dwell times of 20~ms were used after each energy jump and data were taken only during the last 10~ms of each dwell time interval. The data that are shown in Figs.~\ref{fig:allfits}e and  \ref{fig:allfits}f are the data that were accumulated while the electron-ion collision energy had been set to 0.4~eV.

The measurements of data set B were carried out with practically the same ion velocities as were used for data set A. However, the slope of the data set B decay curves (Fig.~\ref{fig:allfits}e and \ref{fig:allfits}f) is steeper than the slope of the data set A curves (Fig.~\ref{fig:allfits}a and \ref{fig:allfits}b). This change in beam lifetime by about 50\% is largely due to differences between the TSR vacuum conditions of October 2010 and those of April 2011.

\subsubsection{Data set C}

For data set C (Figs.~\ref{fig:allfits}i and \ref{fig:allfits}j) a higher ion energy of about 225 MeV was used. The higher energy was realized by operating TSR in synchrotron mode after each injection of a new ion pulse. In order to keep the ions circulating during acceleration the magnetic fields were ramped synchronously to higher values. No more than 50\% of the ions were lost during this phase which lasted 6 s. Ion currents after ramping amounted to up to 17~$\mu$A. The magnetic field in the TSR dipole magnets was increased roughly by a factor of 2 (Tab.~\ref{tab:exppar}). The same sequence of jumps in electron-ion collision energy was applied as for data set B. Because of the higher ion velocities this had to be realized at correspondingly higher laboratory electron energies.

\section{Data analysis and results}

\subsection{Time evolution of state populations}\label{sec:time}

Under the assumption that the target electron density and the partial pressures of the residual-gas components do not change during the measurement, the time evolution of the populations $N_m$ and $N_g$ of the metastable $2s\,2p\;^3P_0$ state and of the $2s^2\;^1S_0$ ground state, respectively, can be obtained from solving the following set of rate equations \cite{Schmidt1994}:
\begin{eqnarray}
\frac{dN_m}{dt} &=& -(A_r + A_q + A_\mathrm{SE} + A_\mathrm{DR} + A_{ml})N_m\nonumber\\ &&+ A_e N_g,\label{eq:dNm}\\
\label{eq:dNg}\frac{dN_g}{dt} &=& (A_r + A_q + A_\mathrm{SE}) N_m - (A_e+A_{gl})N_g,
\end{eqnarray}
where the rates $A_X$ refer to various atomic processes that populate or depopulate states $m$ and $g$. The rate $A_r$ is the radiative HFI transition rate from $m$ to $g$ which will eventually be obtained from the present data analysis. A variety of collision processes can lead to the quenching of $m$ with simultaneous population increase of $g$. Among these processes are collisions of the stored ions with residual gas particles and quenching by external electromagnetic fields. In the above rate equations these processes are accounted for globally by the rate $A_q$. Radiative or dielectronic capture of a free electron by an initially metastable ions followed by Auger decay to the ground state, i.e., superelastic scattering (rate $A_\mathrm{SE}$)
also leads to population transfer from $m$ to $g$. Collisions may also result in excitation from $g$ to $m$ (rate $A_e$) and in loss of ground-state and metastable ions from the storage ring (rates $A_{gl}$ and $A_{ml}$, respectively). Finally, the rate $A_\mathrm{DR}$ accounts for the depopulation of the metastable state by dielectronic recombination. At the experimental electron-ion collision energy this loss process is only available for metastable ions and not for ground-state ions.

With the initial conditions $N_m(0) = N_m^{(0)}$, $N_g(0) = N_g^{(0)}$ and with the definitions

\begin{eqnarray}
 \label{eq:Sigma}\mathcal{A} &=& \frac{1}{2}\left(A_r + A_q + A_\mathrm{SE} + A_e + A_\mathrm{DR} + A_{ml} + A_{gl}\right)\\
 \label{eq:Upsilon}\Upsilon &=& \sqrt{\left(\mathcal{A}-A_{gl}\right)^2-A_e (A_\mathrm{DR} +A_{ml} -A_{gl})},
\end{eqnarray}

the solutions of Eqs.~\ref{eq:dNm} and \ref{eq:dNg} are \cite{Mathematica8}:

\begin{widetext}
\begin{eqnarray}
 \label{eq:Nm}N_m(t) &=& \left\{N_m^{(0)}\cosh(\Upsilon t)+\left[A_e N_g^{(0)}-(\mathcal{A}-A_e-A_{gl})N_m^{(0)}\right]\frac{\sinh(\Upsilon t)}{\Upsilon}\right\}\exp(-\mathcal{A} t),\\
 \label{eq:Ng}N_g(t) &=& \left\{N_g^{(0)}\cosh(\Upsilon t)+\left[(A_r+A_q+A_\mathrm{SE})N_m^{(0)}+(\mathcal{A}-A_e-A_{gl})N_g^{(0)}\right]\frac{\sinh(\Upsilon t)}{\Upsilon}\right\}\exp(-\mathcal{A} t).
\end{eqnarray}
\end{widetext}

Rewriting the hyperbolic functions as sums of two exponentials reveals that the time evolution of the number of stored ions for both states is determined by the two time constants
\begin{eqnarray}
\label{eq:taum}   1/\tau_m &=& \mathcal{A}+\Upsilon,\\
\label{eq:taug}   1/\tau_g &=& \mathcal{A}-\Upsilon.
\end{eqnarray}
which directly reflect the effective storage lifetimes of the metastable and ground-state ions, respectively, as suggested by the indices. As seen earlier \cite{Schmidt1994}, rather simple expressions for these time constants, are obtained in the limit $A_e=0$. In this case, $\Upsilon = \mathcal{A}-A_{gl}$ (Eq.~\ref{eq:Upsilon}), and Eqs.\ \ref{eq:taum} and \ref{eq:taug} yield
\begin{eqnarray}
\label{eq:taum0}   1/\tau_m &=& A_r+A_q+A_\mathrm{SE}+A_\mathrm{DR}+A_{ml}\\
\label{eq:taug0}   1/\tau_g &=& A_{gl}
\end{eqnarray}

As discussed in Sec.~\ref{sec:Ahfi}, this approximation is appropriate in our case. Moreover, it will be shown that further relations between the parameters in these equations can be expected to be approximately fulfilled, so that the radiative lifetime $A_r$ can be found once the decay times $\tau_m$ and $\tau_g$ have been experimentally determined.

In the experiment, the measured count rate of recombined ions comprises contributions from both the metastable and the ground state. Accordingly, it can be represented as the sum
\begin{equation}
\label{eq:Rate1} R(t) = -p_m \frac{dN_m(t)}{dt} - p_g \frac{dN_g(t)}{dt}
\end{equation}
where $p_m$ ($p_g$) is the probability that a metastable (ground state) ion which is lost from the storage ring will then be collected by the recombination detector.  This probability is effectively the rate of capturing an electron either from electron-ion recombination in the Cooler or from the residual gas in the straight section of the storage ring leading up to the detector relative to the rate of all loss processes along the entire ring.

Since the hyperbolic functions in Eqs.~\ref{eq:Nm} and \ref{eq:Ng} can be expressed as the sums of two exponentials, the measured count rate (Eq.~\ref{eq:Rate1}) can also be written as
\begin{equation}\label{eq:Rate2}
 R(t) = c_m e^{\displaystyle  -t/\tau_m}+ c_g e^{\displaystyle -t/\tau_g}
\end{equation}
where the coefficients
\begin{eqnarray}
\label{eq:cm}    c_m &=& \frac{p_m}{2\tau_m}\left[\left(1+\xi\right)N_m^{(0)}-\frac{A_e}{\Upsilon}N_g^{(0)}\right]\\ &&+
                         \frac{p_g}{2\tau_m}\left[\left(1-\xi\right)N_g^{(0)}-\frac{A_r+A_q+A_\mathrm{SE}}{\Upsilon}N_m^{(0)}\right],\nonumber\\
\label{eq:cg}    c_g &=& \frac{p_m}{2\tau_g}\left[\left(1-\xi\right)N_m^{(0)}+\frac{A_e}{\Upsilon}N_g^{(0)}\right]\\ &&+
                         \frac{p_g}{2\tau_g}\left[\left(1+\xi\right)N_g^{(0)}+\frac{A_r+A_q+A_\mathrm{SE}}{\Upsilon}N_m^{(0)}\right],\nonumber
\end{eqnarray}
with
\begin{equation}\label{eq:chi}
    \xi = \frac{\mathcal{A}-A_e-A_{gl}}{\Upsilon}\; \xrightarrow[A_e\to\,0]{}1,
\end{equation}
are obtained by inserting Eqs.\ \ref{eq:Nm} and \ref{eq:Ng} into  Eq.~\ref{eq:Rate1} and comparing the resulting expression with Eq.~\ref{eq:Rate2}.
For the lifetime determination described below, the coefficients $c_m$ and $c_g$, obtained from fitting Eq.~\ref{eq:Rate2} to the data, will not be further considered.

\subsection{Fits to measured decay curves}

\begin{table*}[ttt]
\caption{\label{tab:fit}Isotope-dependent results for the decay times $\tau_{m,g}$ and relative weights $c_{m,g}$ obtained from the fits of Eq.\ \ref{eq:Rate2} to the experimental decay curves displayed in Fig.~\ref{fig:allfits}. The standard deviations for these quantities were obtained from joint confidence intervals for $\tau_m$ and $\tau_g$ and for $c_m$ and $c_g$, respectively (see text). Also given are the mass number $A$ and the reduced $\chi^2$ values of the fits as well as
the derived electron collision rates $A_\mathrm{EC}$ (Eqs.~\ref{eq:AEC32} and \ref{eq:AEC33}) and HFI lifetimes $\tau_\mathrm{HFI}$ (Eq.~\ref{eq:Ahfi}).}
\begin{ruledtabular}
\begin{tabular}{ccr @{\hspace{\fill}} d@{$\pm$\hspace{-3ex}}l @{\hspace{\fill}} d@{\!\!$\pm$\hspace{-3ex}}l @{\hspace{\fill}} r@{$\;\pm$\hspace{-3.5ex}}r @{\hspace{\fill}} r@{$\;\pm$\hspace{-3.5ex}}r @{\hspace{\fill}} r@{$\;\pm$\hspace{-3.5ex}}r @{\hspace{\fill}} c}
data set & $A$ & ~~~$\chi^2$ & \multicolumn{2}{c}{$\tau_m$ (s)} &
                             \multicolumn{2}{c}{$\tau_g$ (s)} &
                             \multicolumn{2}{c}{$c_{m}$ (s$^{-1}$)}  &
                             \multicolumn{2}{c}{$c_{g}$ (s$^{-1}$)}  &
                             \multicolumn{2}{c}{$A_\mathrm{EC}$ (s$^{-1}$)} &
                             $\tau_\mathrm{HFI}$ (s)   \\
                                   \hline\rule[0mm]{0mm}{4mm}
\multirow{2}{*}{{\normalsize A}} & 32 & 1.010 & 33.0 & 1.5    &  136.3  & 0.4           & 451  &  \phantom{1}14   & 3890  & \phantom{4}17  & 0.023 & 0.001 &      \multirow{2}{*}{10.5 $\pm$ 0.5}  \\
                                 & 33 & 1.013 &  8.0 & 0.3    &  122.7  & 0.1           & 722  &  \phantom{1}34   & 6565  & \phantom{14}3  & 0.022 & 0.001&        \\
                                   \hline\rule[0mm]{0mm}{4mm}
\multirow{2}{*}{{\normalsize B}} & 32 & 1.029 & 24.8 &  2.0   &   89.8  & 0.2           & 2226 &             139  & 49024 &           181  & 0.029  &  0.003 & \multirow{2}{*}{10.5 $\pm$ 1.2}  \\
                                 & 33 & 1.026 &  7.4 &  0.6   &   83.4  & 0.1           & 1202 &             130  & 14363 & \phantom{1}10  & 0.028  & 0.003   & \\
                                   \hline\rule[0mm]{0mm}{4mm}
\multirow{2}{*}{{\normalsize C}} & 32 & 1.042 & 30   &  2     &  148    & \phantom{1}6  &  62  &  \phantom{16}2   & 63    & \phantom{14}3  & 0.027   & 0.002 &  \multirow{2}{*}{10.2 $\pm$ 0.7}  \\
                                 & 33 & 0.982 &  7.8 &  0.4   &  177    & \phantom{1}1  & 100  &  \phantom{16}1   & 84    & \phantom{14}1  & 0.025   & 0.002   & \\
\end{tabular}
\end{ruledtabular}
\end{table*}

\begin{figure}
\centering{\includegraphics[width=\figurewidth]{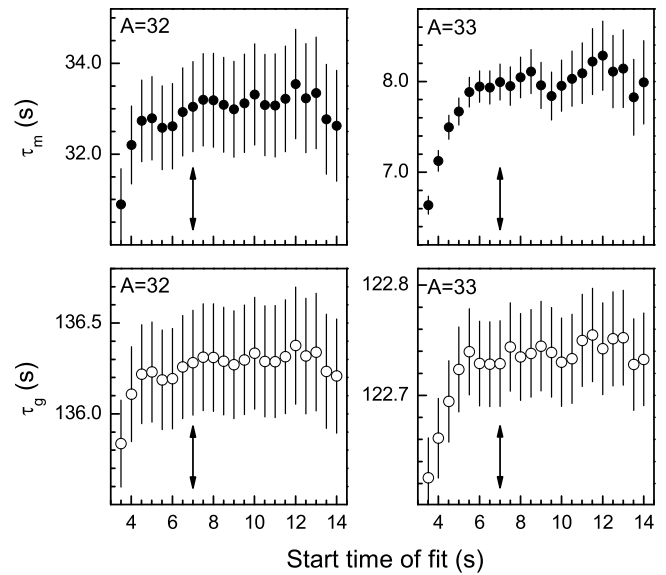}}%
\caption{\label{fig:fitstartA} Dependence of the fit parameters $\tau_m$ (full symbols) and $\tau_g$ (open symbols) on the time of the first data point that is used in the fit of Eq.~\ref{eq:Rate2} to the measured decay curves of data set A (Fig.~\ref{fig:allfits}a and \ref{fig:allfits}b). The arrows mark the data that were used in the calculation of $\tau_\mathrm{HFI}$. (Note that all values shown here scatter by less than the one-sigma error bars since they are statistically not independent, see text.)}
\end{figure}

The results from the fits of Eq.~\ref{eq:Rate2} to the measured decay curves, i.e., the fit curves and the corresponding residuals, are included in Fig.~\ref{fig:allfits}. The best-fit values of the  parameters $\tau_m$, $\tau_g$, $c_m$ and $c_g$ are given in columns 1--7 of Tab.~\ref{tab:fit} along with their standard deviations. A range of initial data points were excluded from the fits in order to suppress the time intervals when the equilibrium conditions of the electron-cooled ion beam were not yet reached. The evolution of the best fit parameters when increasingly more early-data points are excluded from the fits is shown in Figure~\ref{fig:fitstartA}. For data set A, stable conditions were reached only after about 7~s (8~s for data sets B and C). Hence, data from the first 7~s (8~s) after ion injection into the TSR were disregarded in the fits to the three data sets, respectively.

\begin{figure}[t]
\centering{\includegraphics[width=\figurewidth]{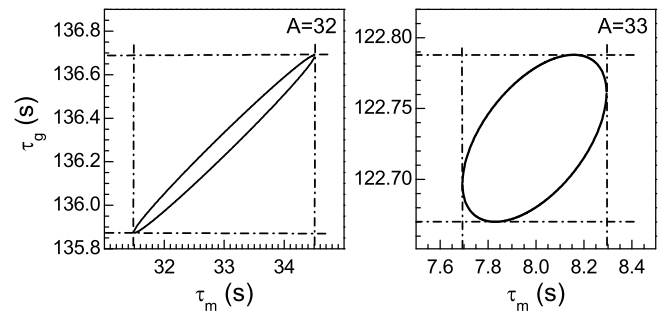}}%
\caption{\label{fig:SIM33ellips} Correlations between the fit parameters $\tau_m$ and $\tau_g$ for data set A  from  the covariance matrices of the least squares fits to the experimental resonance-decay curve (Figs.~\ref{fig:allfits}a and \ref{fig:allfits}b). The ellipses denote the one sigma confidence limits for both $\tau_m$ and $\tau_g$ jointly. The dash-dotted horizontal and vertical lines represent the projections of the ellipses onto the parameter axes.}
\end{figure}

The error bars in Fig.~\ref{fig:fitstartA} correspond to one standard deviation and were obtained from the nonlinear least-squares fitting procedure. These errors neglect mutual correlations between the fit parameters. The correlations can be evaluated from the covariance matrix of the fit \cite{Press2002a} and be visualized as ellipses in two-dimensional subspaces of the fit-parameter space (Fig.~\ref{fig:SIM33ellips}a). In such plots the uncertainties of the individual fit parameters correspond to the projections of the ellipses onto the parameter axes. The resulting uncertainties are the ones that are given in Tab.~\ref{tab:fit}. Since they stem from  confidence regions for two parameters jointly, they are larger (in the present case by 50\%) than the error bars in Fig.~\ref{fig:fitstartA} that represent one-dimensional one-sigma confidence intervals.

\subsection{Extraction of the hyperfine-induced decay rate}\label{sec:Ahfi}

In general, the set of equations \ref{eq:Rate2}, \ref{eq:cm}, and \ref{eq:cg} cannot be used to unambiguously derive all quantities of interest from the fits to the measured decay curves. However, the interpretation of the fitted decay constants in terms of atomic
transition rates is facilitated if the effect of excitation of ground-state ions to the metastable state (rate $A_e$) can be neglected, so that Eqs.~\ref{eq:taum0} and \ref{eq:taug0} can be applied. Collisional excitation
has been investigated in previous experimental measurements of metastable
state lifetimes, e.\,g., with stored C$^{4+}$ \cite{Schmidt1994}, Xe$^+$ \cite{Mannervik1997a},
and Li$^+$ \cite{Saghiri1999} ions. In all
these cases collisional processes were found to have a negligibly small
influence on the population of the metastable states under investigation,
mainly because of the low residual particle density in the ultrahigh
vacuum of the storage ring. Since the residual gas pressure and the
collisional excitation cross sections of the present experiment and of
the above mentioned studies are of the same order of magnitude,
collisional excitation processes are assumed to be
negligible within the experimental uncertainty, and, accordingly, Eqs.~\ref{eq:taum0} and \ref{eq:taug0} are considered to hold.

Deexcitation processes in the residual gas, denoted by $A_q$, are neglected here based on the same argument used above for the collisional excitation processes ($A_e$) and we set $A_q=0$. Since the rate $A_q$  also accounts for the quenching of the metastable level by external magnetic and electric fields these effects are implicity also assumed to be negligible. This assumption is justified in view of theoretical investigations of Stark quenching of the $2s\,2p\;^3P_0$ level in Be-like ions \cite{Maul1998a} and of magnetic quenching $2s\,2p\;^3P_0$ level in the Be-like  Ti$^{18+}$ \cite{Li2011}. Accordingly, quenching by electric fields  \cite{Maul1998a} becomes noticeable only at field strengths which are orders of magnitude larger than the motional electric fields in the TSR dipole magnets. Likewise, in Ref.~\cite{Li2011} it was shown that magnetic fields with strengths that are typically encountered in the TSR do not significantly influence the lifetime of the Ti$^{18+}$($2s\,2p\;^3P_0$) level. The present measurements which were performed at different settings of the TSR dipole magnets
confirm this theoretical finding (see discussion below).

With $A_q=0$, Eq.~\ref{eq:taum0} reads
\begin{equation}\label{eq:taum1}
    \frac{1}{\tau_m} = A_r +  A_{ml} + A_\mathrm{DR} + A_\mathrm{SE}.
\end{equation}
For finding $A_r$ the rates $A_{ml}$, $A_{\rm DC}$ and $A_\mathrm{SE}$ still need to be specified. In order to provide a value for the rate $A_{ml}$ we make use of the plausible assumption that the collisional rates of the ground and the metastable state are essentially equal, i.e., that $A_{ml}=A_{gl}$.  This assumption is based on the fact that the ionization energies of the $^1S_0$ ground state and the $^3P_0$ metastable state, 652.2~eV and 627.5~eV, respectively \cite{Ralchenko2011}, differ by less than 4\%. It does not lead to any serious consequences. When, e.\,g., $A^{(A)}_{ml}=2 A^{(A)}_{gl}$ is assumed, the final result for the HFI transition rate changes by less than 1.5\%. This change is smaller than the 4\% experimental uncertainty (see below).

Inserting $A_{ml}=A_{gl}$ and $A_{gl}=1/\tau_{g}$ (Eq.~\ref{eq:taug0}) into Eq.~\ref{eq:taum1} yields
\begin{equation}\label{eq:Art}
    A_r = \frac{1}{\tau_m} - \frac{1}{\tau_g} - A_\mathrm{EC}
\end{equation}
where the electron collision (EC) rate \footnote{In Ref.~\cite{Schippers2007a} the rate $A_\mathrm{EC}$ is denoted as $A_\mathrm{DC}$.} $A_\mathrm{EC} =  A_\mathrm{DR}+ A_\mathrm{SE}$ accounts for dielectronic recombination of and superelastic scattering from initially metastable ions.  Exploiting the fact that there is no HFI transition for the $A=32$ isotope, i.e., $A_r^{(32)} = 0$, Eq.~\ref{eq:Art} yields the EC rate for this isotope:
\begin{equation}\label{eq:AEC32}
    A_\mathrm{EC}^{(32)} = \frac{1}{\tau^\mathrm{(32)}_m} - \frac{1}{\tau^\mathrm{(32)}_g}.
\end{equation}
The electron collision rate can be expressed as $A^{(A)}_\mathrm{EC} = n^{(A)}_e \alpha_\mathrm{EC}$ where the rate coefficient $\alpha_\mathrm{EC}$ is the same for the metastable ions of both isotopes. Consequently, the EC rate for the $A=33$ isotope is
\begin{equation}\label{eq:AEC33}
    A_\mathrm{EC}^{(33)} = \frac{n_e^{(33)}}{n_e^{(32)}}A_\mathrm{EC}^{(32)}=\frac{n_e^{(33)}}{n_e^{(32)}}\left(\frac{1}{\tau^\mathrm{(32)}_m} - \frac{1}{\tau^\mathrm{(32)}_g}\right).
\end{equation}
In the previous experiment with Be-like Ti$^{18+}$ \cite{Schippers2007a} the density ratio was very close to 1 and therefore omitted from the equations. In the present measurements the ratios were 0.938, 0.948, and 0.932 for data sets A, B, and C, respectively (Tab.~\ref{tab:exppar}). As can be seen from the next to last column of Tab.~\ref{tab:fit} the rates $A_\mathrm{EC}$  for the three experimental data sets mutually agree within 5\%.

Using now Eq.~\ref{eq:Art} for the $A=33$ isotope gives the $2s\,2p\;^3P_0\to 2s^2\;^1S_0$ transition rate in the laboratory frame:
\begin{equation}
 A_r^{(33)} = \frac{1}{\tau_m^{(33)}}-\frac{1}{\tau_g^{(33)}}-\frac{n_e^{(33)}}{n_e^{(32)}}\left(
     \frac{1}{\tau_m^{(32)}}-\frac{1}{\tau_g^{(32)}}\right).
\end{equation}
Finally, the HFI transition rate is obtained from a transformation into the reference frame of the moving ions, i.e., by multiplication of $A_r^{(33)}$ with
the Lorentz factor $\gamma^{(33)} = [1-(v_\mathrm{ion}^{(33)}/c)^2]^{-1/2}$ ($c$ denotes the speed of light):
\begin{eqnarray}\label{eq:Ahfi}
    A_\mathrm{HFI} &=& \frac{1}{\tau_\mathrm{HFI}} \\
    &=& \gamma^{(33)}\left[\frac{1}{\tau_m^{(33)}}-\frac{1}{\tau_g^{(33)}}-\frac{n_e^{(33)}}{n_e^{(32)}}\left(
     \frac{1}{\tau_m^{(32)}}-\frac{1}{\tau_g^{(32)}}\right)\right].\nonumber
\end{eqnarray}

\begin{table}[b]
\caption{\label{tab:s12comp}Comparison of theoretical values for the hyperfine induced $2s\,2p\; ^3P_0 \to 2s^2\; ^1S_0$ transition rate in Be-like S$^{12+}$ with the present experimental result.}
\begin{ruledtabular}
\begin{tabular*}{\columnwidth}{@{\extracolsep{\fill}}ccccc}
\multirow{2}{*}{Year} & \multicolumn{2}{c}{$A_\mathrm{HFI}$ (s$^{-1}$)} & Deviation from & \multirow{2}{*}{Ref.} \\
     & Experiment & Theory & experiment & \\
     \hline\rule[0mm]{0mm}{4mm}%
 1993 &            & 0.03611 & -62.4\%               & \cite{Marques1993} \\
 2008 &            & 0.09315 & -\phantom{6}3.0\%     & \cite{Cheng2008a}   \\
 2009 &            & 0.09355 & -\phantom{6}2.5\%     & \cite{Andersson2009}\\
 2011 &  0.096(4)  &  & &  present\\
\end{tabular*}
\end{ruledtabular}
\end{table}

Results for the hyperfine induced lifetime as derived from the three experimental data sets A, B, and C, are given in the last column of Tab.~\ref{tab:fit}. This table also lists the best fit values for the decay time constants which were used in Eq.~\ref{eq:Ahfi} together with the experimental parameters from Tab.~\ref{tab:exppar}. Although the three data sets were obtained under significantly different experimental conditions, the three experimental values for $\tau_\mathrm{HFI}$ agree with each other within the experimental uncertainties. This documents the robustness of the present experimental approach. The weighted mean (with the inverse squared errors as weights) of the three experimental HFI lifetimes is $\tau_\mathrm{HFI} = 10.4\pm0.4$~s.

\section{Discussion}\label{sec:discussion}

During the measurement of data set C the magnetic field in the TSR bending magnets was a factor of two stronger as compared to the measurements of data sets A and B (Tab.~\ref{tab:exppar}). At the same time, $\tau_\mathrm{HFI}$ from data set C is smaller by 3\% than $\tau_\mathrm{HFI}$ from data sets A and B (Tab.~\ref{tab:fit}). This may be an indication of a slight quenching of the $2s\,2p\;^3P_0$ state in the magnetic fields of the storage ring magnets (see also \cite{Mannervik1996a}) which cover 13\% of the closed orbit of the stored ions (Fig.~\ref{fig:TSR}). This effect is expected to increase linearly with increasing magnetic field strength. Since the 3\% change of $\tau_\mathrm{HFI}$ is smaller than the 4\% experimental uncertainty, it must, however, be concluded that there is no significant influence of the TSR magnetic fields on the measured HFI transition rates at the present level of experimental accuracy. This conclusion is in accord with recent theoretical results for isoelectronic Ti$^{18+}$ \cite{Li2011}.

In Tab.~\ref{tab:s12comp} the present experimental result for the HFI transition rate $A_\mathrm{HFI} = 1/\tau_\mathrm{HFI} = 1/10.4(4)$~s$^{-1} = 0.096(4)$~s$^{-1}$ is compared with the available theoretical results \cite{Marques1993,Cheng2008a,Andersson2009}. The early theoretical calculation by Marques et al.~\cite{Marques1993} yielded a value that is about 60\% lower than the experimentally determined rate. This is in contrast to
the later calculations by Cheng et al.\ \cite{Cheng2008a} and  Andersson et al.\ \cite{Andersson2009} which both yield results in very good agreement with the present experimental value. As already discussed (Sec.~\ref{sec:intro}), the main difference between the early and later calculations is the treatment of electron-electron correlation effects. In particular, Marques et al.~\cite{Marques1993} disregarded the mixing of the $2s\,2p\;^1P_1$ level with the $2s\,2p\;^3P_1$ level. This mixing was included in the later calculations of HFI transition rates in Be-like ions \cite{Brage1998a,Cheng2008a,Andersson2009,Li2010}. The comparison of the available theoretical results with the present experimental result for $^{33}$S$^{12+}$ strongly suggests that the additional computational effort of the later calculations is indeed required for arriving at more accurate HFI transition rates.

Generally, the relative importance of correlation effects decreases with charge state. In the present context, this can be seen from the differences between the early and the later theoretical HFI transition rates. For the lower charged $^{33}$S$^{12+}$ ion the result of Marques et al.\ \cite{Marques1993} deviates by 61\% from the result of Cheng et al.~\cite{Cheng2008a} (Tab.~\ref{tab:s12comp}). For the more highly charged $^{47}$Ti$^{18+}$ ion this difference is only 47\%  (Tab.~\ref{tab:ti18comp}). In view of this fact, the $\sim$20\% disagreement of the experimental result for $^{47}$Ti$^{18+}$ with the latest theoretical calculation \cite{Cheng2008a,Andersson2009}, which agree well with the $^{33}$S$^{12+}$ experimental value, is surprising. In the attempt to find an explanation for this disturbing discrepancy, the influence of external magnetic fields on the lifetime of the $^{47}$Ti$^{18+}$($2s\,2p\;^3P_0$) has been investigated by detailed theoretical calculations \cite{Li2011}. From this study it was concluded that the magnetic fields in the the TSR bending magnets do not influence the measured HFI lifetimes significantly. This is in agreement with the present experimental findings for $^{33}$S$^{12+}$ ions.

\section{Conclusions}

We have measured the hyperfine induced $2s\,2p\;^3P_0\to2s^2\;^1S_0$ transition rate  in Be-like $^{33}$S$^{12+}$ ions using isotopically pure ion beams in a heavy-ion storage ring. The decay of the number of stored ions was monitored by employing electron-ion recombination. In order to selectively enhance the signal from the $2s\,2p\;^3P_0$ level, the electron-ion collision energy was tuned to a dielectronic recombination resonance associated with this level.

In our present experimental work, different data sets were obtained under substantially different conditions. Nevertheless, the individual HFI transition rates from the different data sets agree with each other within the experimental uncertainties. This demonstrates the robustness of the present experimental approach.

Our experimental value of 0.096(4)~s$^{-1}$ agrees with the latest theoretical calculations \cite{Cheng2008a,Andersson2009}. No significant influence of the magnetic field in the storage-ring bending magnets on our result has been found. The origin of the discrepancy between our previously measured HFI rate for $^{47}$Ti$^{18}$ \cite{Schippers2007a} and the latest calculations \cite{Cheng2008a,Andersson2009,Li2010} still remains unclear.

\begin{acknowledgments}
We thank  the MPIK accelerator and TSR crews for their excellent support. Financial support by the Deutsche Forschungsgemeinschaft (DFG, grant no.\ SCHI378/8-1) and by the Max-Planck Society  is gratefully acknowledged. MH, ON, and DWS were financially supported in part by the NASA Astronomy and Physics Research and Analysis program and the NASA Solar Heliospheric Physics program.
\end{acknowledgments}

\end{document}